\documentclass{article}

\usepackage{arxiv}

\usepackage[utf8]{inputenc} % allow utf-8 input
\usepackage[T1]{fontenc}    % use 8-bit T1 fonts
\usepackage{hyperref}       % hyperlinks
\usepackage{url}            % simple URL typesetting
\usepackage{booktabs}       % professional-quality tables
\usepackage{amsfonts}       % blackboard math symbols
\usepackage{nicefrac}       % compact symbols for 1/2, etc.
\usepackage{microtype}      % microtypography
\usepackage{lipsum}
\usepackage{graphicx}
\graphicspath{ {./images/} }

\title{Accuracy Enhancement Method for Speech Emotion Recognition from Spectrogram using Temporal Frequency Correlation and Positional Information Learning through Knowledge Transfer}

\author{
 Jeong-Yoon Kim \\
  Department of Electronic Engineering\\
  Hanbat National University\\
  Daejeon, 34158 \\
  \texttt{kjy7567@o365.hanbat.ac.kr} \\
   \And
 Seung-Ho Lee \\
  Department of Electronic Engineering\\
  Hanbat National University\\
  Daejeon, 34158 \\
  \texttt{shlee@cad.hanbat.ac.kr} \\
}
\date{JAN}
\begin{document}
\maketitle
\begin{abstract}
In this paper, we propose a method to improve the accuracy of speech emotion recognition (SER) by using vision transformer (ViT) to attend to the correlation of frequency (y-axis) with time (x-axis) in spectrogram and transferring positional information between ViT through knowledge transfer. The proposed method has the following originality i) We use vertically segmented patches of log-Mel spectrogram to analyze the correlation of frequencies over time. This type of patch allows us to correlate the most relevant frequencies for a particular emotion with the time they were uttered. ii) We propose the use of image coordinate encoding, an absolute positional encoding suitable for ViT. By normalizing the x, y coordinates of the image to -1 to 1 and concatenating them to the image, we can effectively provide valid absolute positional information for ViT. iii) Through feature map matching, the locality and location information of the teacher network is effectively transmitted to the student network. Teacher network is a ViT that contains locality of convolutional stem and absolute position information through image coordinate encoding, and student network is a structure that lacks positional encoding in the basic ViT structure. In feature map matching stage, we train through the mean absolute error (L1 loss) to minimize the difference between the feature maps of the two networks. To validate the proposed method, three emotion datasets (SAVEE, EmoDB, and CREMA-D) consisting of speech were converted into log-Mel spectrograms for comparison experiments. The experimental results show that the proposed method significantly outperforms the state-of-the-art methods in terms of weighted accuracy while requiring significantly fewer floating point operations (FLOPs). Moreover, the performance of the student network is better than that of the teacher network, indicating that the introduction of L1 loss solves the overfitting problem. Overall, the proposed method offers an promising solution for SER by providing improved efficiency and performance.
\end{abstract}

% keywords can be removed

\section{Introduction}
Speech-related tasks, such as speech to text (STT) and speech emotion recognition (SER), are becoming increasingly important in our daily lives, with applications in intelligent speakers, voice assistants, and more. Early studies utilized only the magnitude of speech samples\cite{er2020novel},\cite{daneshfar2020speech},\cite{garg2013speech},\cite{hou2020supervised},\cite{tarantino2019self}, but recently, methods used in visual tasks have been introduced to analyze spectrogram, which convert frequency information of speech into images\cite{maji2022advanced},\cite{andayani2022hybrid},\cite{nagarajan2020speech},\cite{chen2021impact},\cite{xu2021head},\cite{meng2019speech},\cite{jiang2019parallelized},\cite{bertero2017first},\cite{gong21b_interspeech},\cite{DBLP:journals/corr/abs-2203-09581}. The most popular method for converting speech samples into spectrogram images is the short-time Fourier transform (STFT). The STFT is a Fourier-related transform used to determine the frequency components of a small receptive field of a time-varying signal. Derives the frequency components from the magnitude of the speech sample using STFT, which are then mapped to the mel scale to more closely match human hearing characteristics, creating a log-Mel spectrogram. In recent speech-related research, the log-Mel spectrogram is used for analysis and classification. 

We focuses on SER among various speech-related tasks, and for this purpose, we compare pros and cons of convolutional neural networks (CNNs)\cite{lecun2015deep} and vision transformer (ViT)\cite{DBLP:conf/iclr/DosovitskiyB0WZ21}, which are commonly used for vision tasks. CNNs are characterized by including and analyzing pixels near the input data and have been used in various fields for a long time. In particular, the locality obtained by analyzing pixels near the input data plays a big role in improving the accuracy of vision tasks. However, CNNs are prone to overfitting due to excessive locality, are not suitable for large datasets, and can lead to inaccurate results when there is a large difference from the input data. To deal with these issues, ViT have emerged as an alternative to CNNs. Generally, ViT segment the input image into multiple square patches and then perform global processing through multi-head self-attention, which has the potential to significantly improve the performance of vision tasks. However, global processing is characterized by slow convergence, need large datasets, and difficult optimization. Furthermore, ViT is sensitive to optimizers, dataset-appropriate learning hyperparameters, training schedules, and network depth, requiring many experiments to empirically select values. In recent study has proposed a method that combines the advantages of CNNs with the advantages of ViT\cite{xiao2021early}. This method replaces ViT's patchify process with several convolutional layers to initially extract local features and then analyze them globally using ViT. The multiple convolutional layers used in the patchify process are called convolutional stem and serve as a robust improvement over ViT's slow convergence speed and variation in learning hyperparameters. However, the application of convolutional stem slightly but clearly increases the number of trainable parameters, which increases the required resources. Also, if the convolutional stem does not perform size reduction of the image and retains the rich information, it will require more resources.

Therefore, in this paper, we propose ViT, which uses knowledge transfer to learn the advantages of the convolution without increasing resources, and vertically partitioned patches to analyze the frequency correlation over time in the log-Mel spectrum. We also propose the use of image coordinate encoding, which is an absolute positional encoding method suitable for ViT, assuming that the locality inference ability of convolutional stem is insufficient. We design a teacher network and a student network for knowledge transfer. Teacher network is a ViT with convolutional stem and image coordinate encoding, and student network is a basic ViT with vertically partitioned patches and no location encoding. Figure 1 shows the overall process of the proposed method in this paper.

The main contributions of this paper are as follows. i: Analyze how the patch shape (square vs. vertically elongated) of ViT can affect SER accuracy by comparing attention masks; ii: Propose the use of image coordinate encoding suitable for ViT; iii: Empirically prove that it is possible to transfer the knowledge of the teacher network to the student network and reproduce it without convolutional stem and positional encoding.

\begin{figure*}[t!]
    \centering
    \includegraphics[width=350pt]{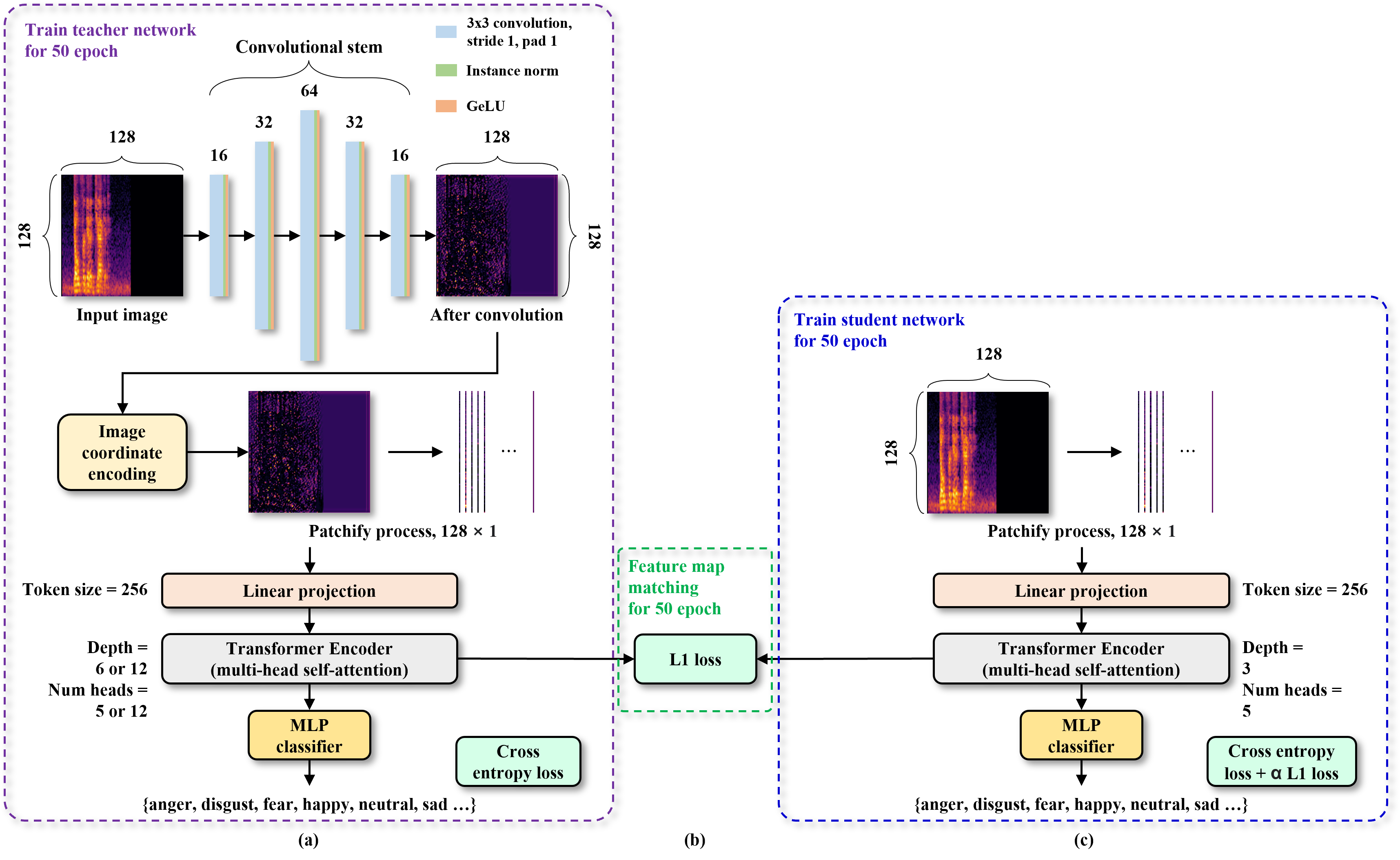}
    \caption{The overall process of proposed method in this paper. i: (a) Train the teacher network using Cross entropy loss. ii: (b) Match the feature maps of the student network and the trained teacher network using Mean absolute error(L1 loss). iii: (c) Train the student network with the feature map matching performed using cross entropy loss + $\alpha$ * L1 loss. Each stage is performed repeatedly with 50 epochs.}
    \label{fig1}
\end{figure*}

\section{Related works}
In the early days of speech signal processing, automatic speech recognition (ASR)\cite{DBLP:conf/interspeech/KandaYGWMCY21},\cite{DBLP:conf/interspeech/LohrenzLF21},\cite{leong2021online} and speech to text (STT)\cite{baevski2020wav2vec} were important due to the development of voice assistant technology to provide hands-free help. However, recent speech signal processing research has placed considerable emphasis on SERs, with efforts to develop more accurate emotion recognition models using log-Mel spectrogram-based features. While early approaches focused primarily on more classical machine learning-based methods, there has been a recent shift toward deep neural network-based models. Among them, methods that utilize the attention mechanism have seen a noticeable increase.

Dosovitskiy et al.\cite{DBLP:conf/iclr/DosovitskiyB0WZ21} proposed ViT, an application of attention-only transformers to image classification. They applied a transformer encoder to image classification and used patches of 16 x 16 segmented input images instead of word token embeddings as the input sequence. ViT performed poorly when trained on small datasets, but excelled when trained on large datasets.  These results suggested that transformers could replace much of the work done by CNNs, often with higher accuracy, and many speech processing methods based on ViT have emerged.

Gong et al.\cite{gong21b_interspeech} proposed an audio spectrogram transformer (AST) that analyzes log-Mel spectrogram by dividing them into square patches using ViT. They further trained on log-Mel spectrogram using DeiT\cite{touvron2021training}, which is a knowledge distillation\cite{hinton2015distilling} from CNNs trained on the ImageNet dataset\cite{deng2009imagenet}. In the patchify process, they partitioned a 16 x 16 patch with 6 pixel overlap to enable fine-grained analysis of log-Mel spectrogram through self-attention.

Ristea et al.\cite{DBLP:journals/corr/abs-2203-09581} proposed a separable transformer (SepTr) that analyzes a spectrogram twice, horizontally and vertically. SepTr consists of a horizontal transformer that analyzes the log-Mel spectrogram in the horizontal direction and a vertical transformer that analyzes it in the vertical direction. In the patchify process, we split the log-Mel spectrogram into 1 x 1 patches to the analysis.

However, transformer-based methods still suffer from the problem that they are sensitive to hyperparameters during training and difficult to optimize. Xiao et al.\cite{xiao2021early} determined that the problem with transformer-based methods is the lack of locality, which can be achieved through hyperparameter-robust and relatively easy-to-optimize CNNs. They replaced ViT's patchify process with a convolutional stem consisting of multiple layers of CNNs to achieve locality while improving on the problems of transformer-based methods. Chen et al.\cite{chenempirical} found that freezing the weights of the patchify process in ViT's self-supervised learning to a random initialization improves stability. They also found that removing positional embeddings only slightly reduced accuracy.

Islam et al.\cite{islam2020much} explored the extent to which CNNs encode spatial positional information and its impact on vision tasks such as semantic segmentation and salient object detection. They found that CNNs inferred information about spatial positioning from zero-padding located at the boundaries of the image and that they relied on and learned from positional information to a much greater extent than expected. In particular, positional information was more prominent in the deeper layers of the CNNs. Baevski et al.\cite{baevski2020wav2vec} used the inference of positional information using these CNNs as a substitute for positional encoding in the transformer, achieving high accuracy for speech recognition using the magnitude of the speech sample.

We determined that for 1D speech magnitude data such as \cite{baevski2020wav2vec} , it is easy to infer absolute positional information from a small number of layers of CNNs. However, for 2D spectrogram, a small number of CNNs is insufficient and absolute positional encoding is required separately at the input of the transformer. Therefore, as a method suitable for ViT, we perform image coordinate encoding by normalizing the coordinates of 2D images to -1 to 1 and concatenating them to the image. In addition, it is common for existing spectrogram analysis methods to use a square-shaped receptive field. However, we assume that it would be more effective to correlate the frequency (y axis) corresponding to a specific emotion with the time (x axis) at which it was uttered, so we use a receptive field that can completely segment the spectrogram vertically.

\section{background}
\subsection{short-time Fourier transform}
We convert each audio sample into a 2D time-frequency matrix to get an image-like representation. To do this, we compute the discrete short-time Fourier transform (STFT) as follows:

\begin{equation}
STFT(m, k)=\sum \limits_{n=-\infty}^{\infty}x[n]\cdot w[n-aH]\cdot e^{-j \frac{2\pi}{N}kn}
\end{equation}

Given input discrete signal $x[n]$, $w[n]$ is a window function (in this paper, Hamming) of length $L, H$ is the hop size, and $N$ is the total number of discrete Fourier transform (DFT) points (frequency bins). $STFT(m, k)$ is STFT coefficient for the $k^{th}$ frequency
bin and the $m^{th}$ time-frame

\subsection{Log-Mel spectrogram}
The Mel scale is a scale of frequencies that a listener determine to be the same distance from each other. The human ear easily distinguish the difference between frequencies up to 1000 Hz, but cannot distinguish the difference between 10 kHz and 10.5 kHz. Therefore, it is converted to Mel scale to match human hearing characteristics. The spectrogram obtained by STFT is log-transformed and Mel-scaled to become a log-Mel spectrogram. The Mel scale is linear up to 1 kHz and logarithmic for frequencies above that. To convert the spectrogram to Mel scale, the computed spectrogram is passed through a Mel-filter bank.

\begin{equation}
Mel(f)=2595\log (1+\frac{f}{700})
\end{equation}

\section{methods}
\subsection{Convolutional stem}
In this paper, we design a convolutional stem consisting of six 3 x 3 convolutional layers to verify that the locality of the convolution can be learn to ViT through knowledge transfer. The output channels of the convolutional stem are [16, 32, 64, 32, 16, 1], respectively, and are output with stride = 1 and zero-padding = 1 to retain the input image size. We retain size to avoid large differences in downsizing and to make the student network and input sequence sizes similar in order to analyze the impact of convolutional stem. Batch normalization is a more common method, but since we use 4 mini-batches, similar to the experiment in \cite{DBLP:journals/corr/abs-2203-09581}, we use instance normalization, which is more advantageous with fewer mini-batches. For the activation function, we use gaussian error linear unit (GELU). GELU is bounded differently than the more common rectified linear unit (ReLU), LeakyReLU, and is therefore more immune to gradient vanishing. Figure 1 shows the structure of the convolutional stem in this paper.

\begin{figure}[h]
    \centering
    \includegraphics[width=350pt]{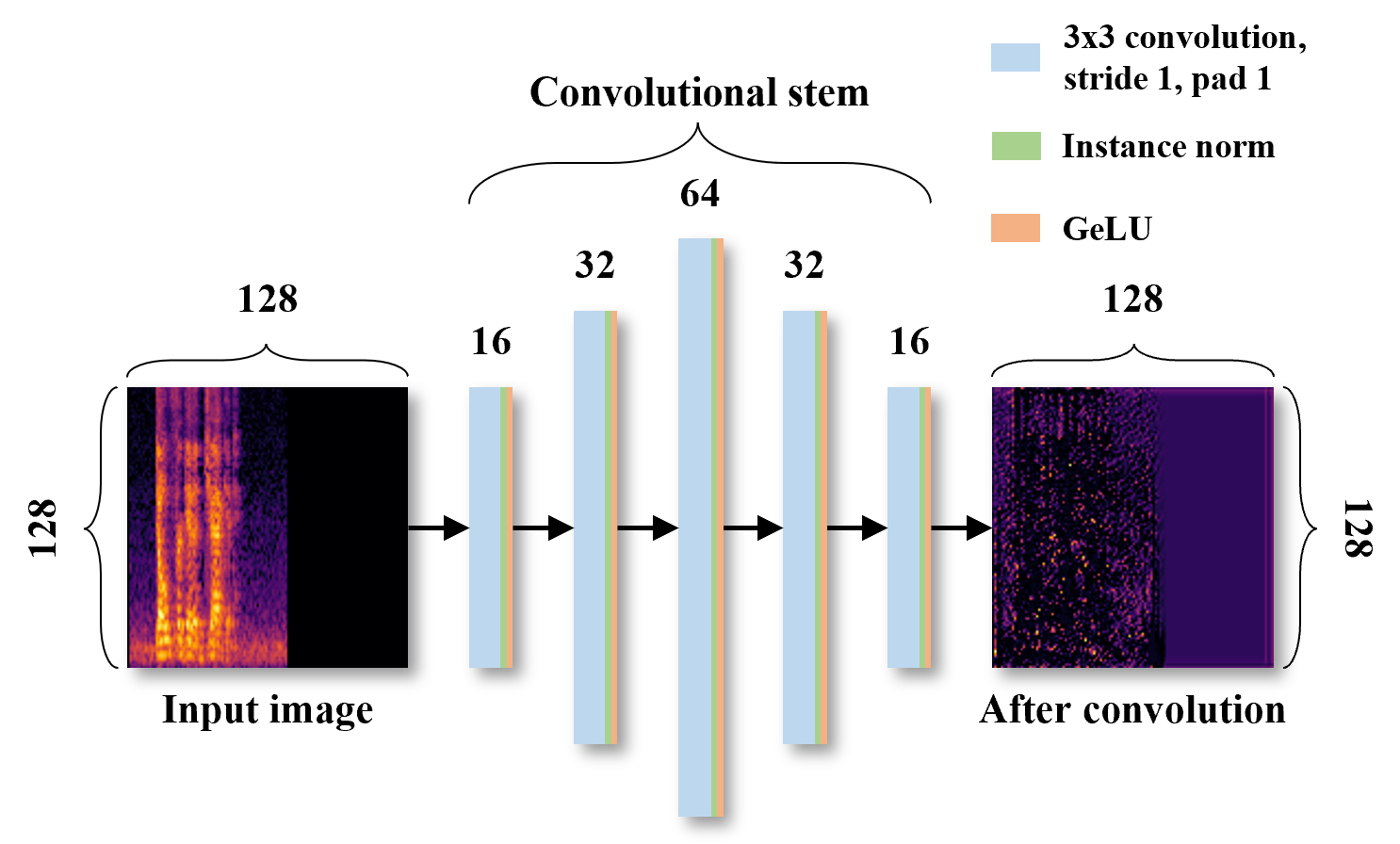}
    \caption{Structure of convolutional stem in this paper. The numbers above each block represent the output channels. To ensure that no information is lost, the after convolution is equal to the input size of 128 x 128.}
    \label{fig2}
\end{figure}

\subsection{Image coordinate encoding}
Relative position encoding, absolute position encoding, and position embedding are methods that are often used in the context of sequence data, such as in natural language processing or computer vision tasks. These methods aim to provide the network with information about the positional relationships between elements. On the other hand, if a transformer is used after a network that can infer positional information, such as CNNs, good performance can be obtained without performing a separate positional encoding[]. However, since only relative positional information can be inferred initially through border recognition through zero-padding, more trains are needed to infer absolute position from this information, convergence is slow, and overfitting occurs before global minima. Therefore, in this paper, image coordinate encoding is performed by normalizing and concatenating the coordinates of log-Mel spectrogram through convolutional stem to values from -1 to 1 (see figure 3).

\begin{figure}[h]
    \centering
    \includegraphics[width=350pt]{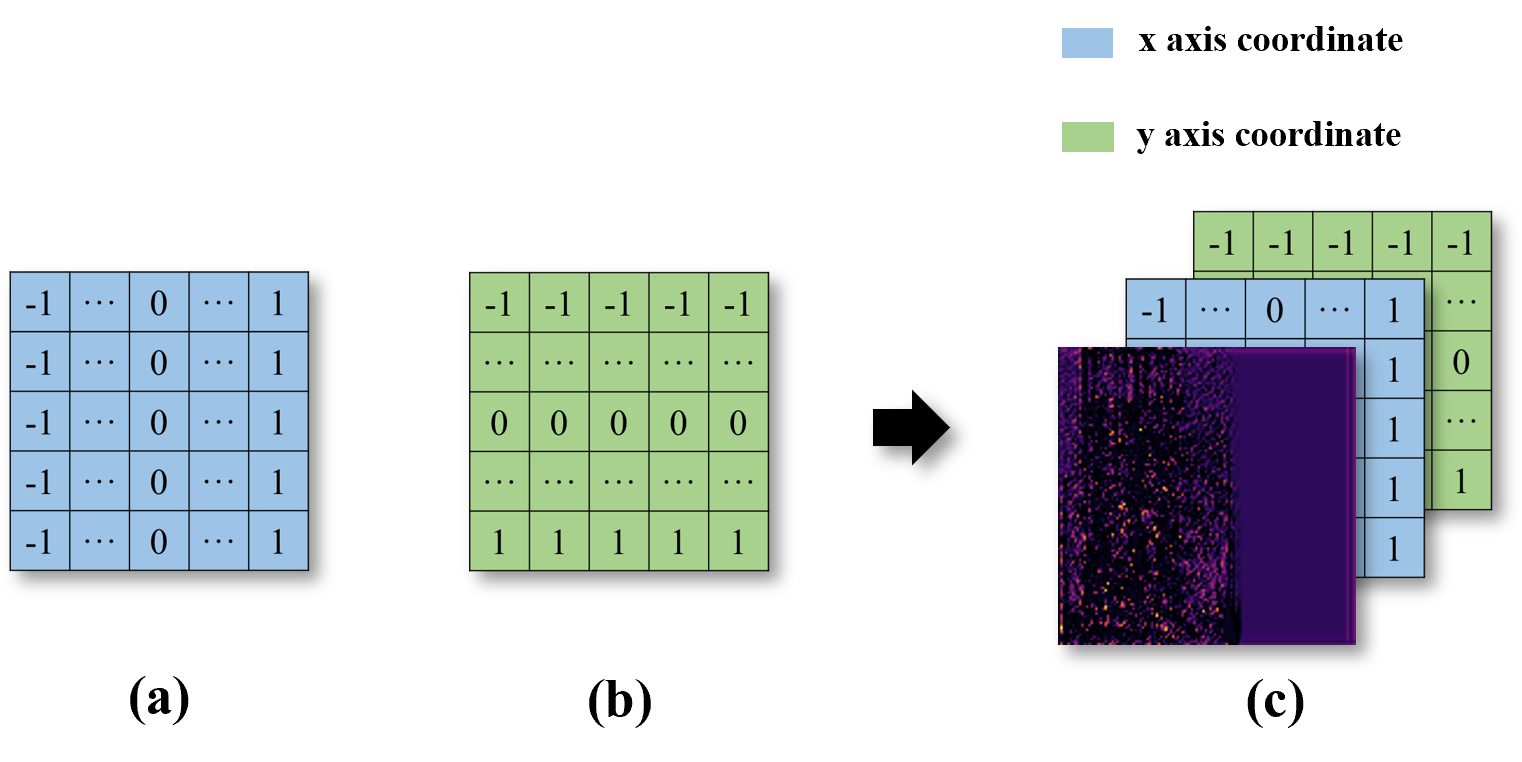}
    \caption{Image coordinate encoding. The x, y coordinates of the image are normalized to the range -1 to 1, (a) and (b) concatenated to a log-Mel spectrogram after convolutional stem, (c).}
    \label{fig3}
\end{figure}

\subsection{Multi-head self-attention}
Transformer's encoder and decoder are both based on the attention mechanism. The attention mechanism is designed to compensate for the weakness of the seq2seq model, which processes each word in a sentence. The seq2seq model can only refer to words near the output layer. With attention, however, all words can be referenced in the calculation of the result without being bound by short- or long-term dependencies. In vision tasks, attention calculates how much attend to each input sequence (i.e., patch) to determine how much each input should be reflected in the resulting computation, i.e., in tasks such as object detection and image classification, more weight is given to important patches when they are segmented from the entire image. Self-attention can learn correlations between input sequences by attend to itself. The self-attention layer consists of three trainable weight matrices that are used to derive a query Q, key K, and value V from an input sequence X. The output attention(Q, K, V) is the weight of the input sequence. The output attention(Q, K, V) as follows:

\begin{equation}
Attention(Q,K,V) = softmax(\frac{Q\cdot K^T}{\sqrt{d_k}})\cdot V
\end{equation}

$K^T$ represents the transpose of K, and $d_k$ represents the dimension of K.
Multi-head self-attention is a module for executing the self-attention mechanism multiple times in parallel. Intuitively, multi-head self-attention allows for more accurate analysis because it can attend to multiple subspaces differently.

\subsection{Network architecture}
In this paper, we design two networks to verify that features with convolutional locality and absolute positional encoding can be implemented through basic ViT without positional encoding. The teacher network with convolutional stem and image coordinate encoding, and the student network without positional encoding in a basic structure ViT are designed for knowledge transfer. Knowledge transfer is performed in the direction of minimizing the difference between the features immediately before each multi-layer perceptron (MLP) classifier.

The teacher network consists of a convolutional stem and image coordinate encoding, ViT with token size = 256, and an MLP classifier. ViT is tuned to depth = 6 or 12 and num heads = 5 or 12 to compare performance with methods such as \cite{gong21b_interspeech}, \cite{DBLP:journals/corr/abs-2203-09581}. In general, the deeper the depth and the more heads, the more difficult it is to optimize for small-sized datasets. However, higher performance can be expected when sufficiently large data is available. Therefore, in this paper, several sizes of ViT are created and used in the experiments for objective comparison. The patch size is 128 x 1 (height, width), which can completely split a 128 x 128 image vertically. Figure 4 shows the structure of the teacher network.

\begin{figure}[t!]
    \centering
    \includegraphics[width=350pt]{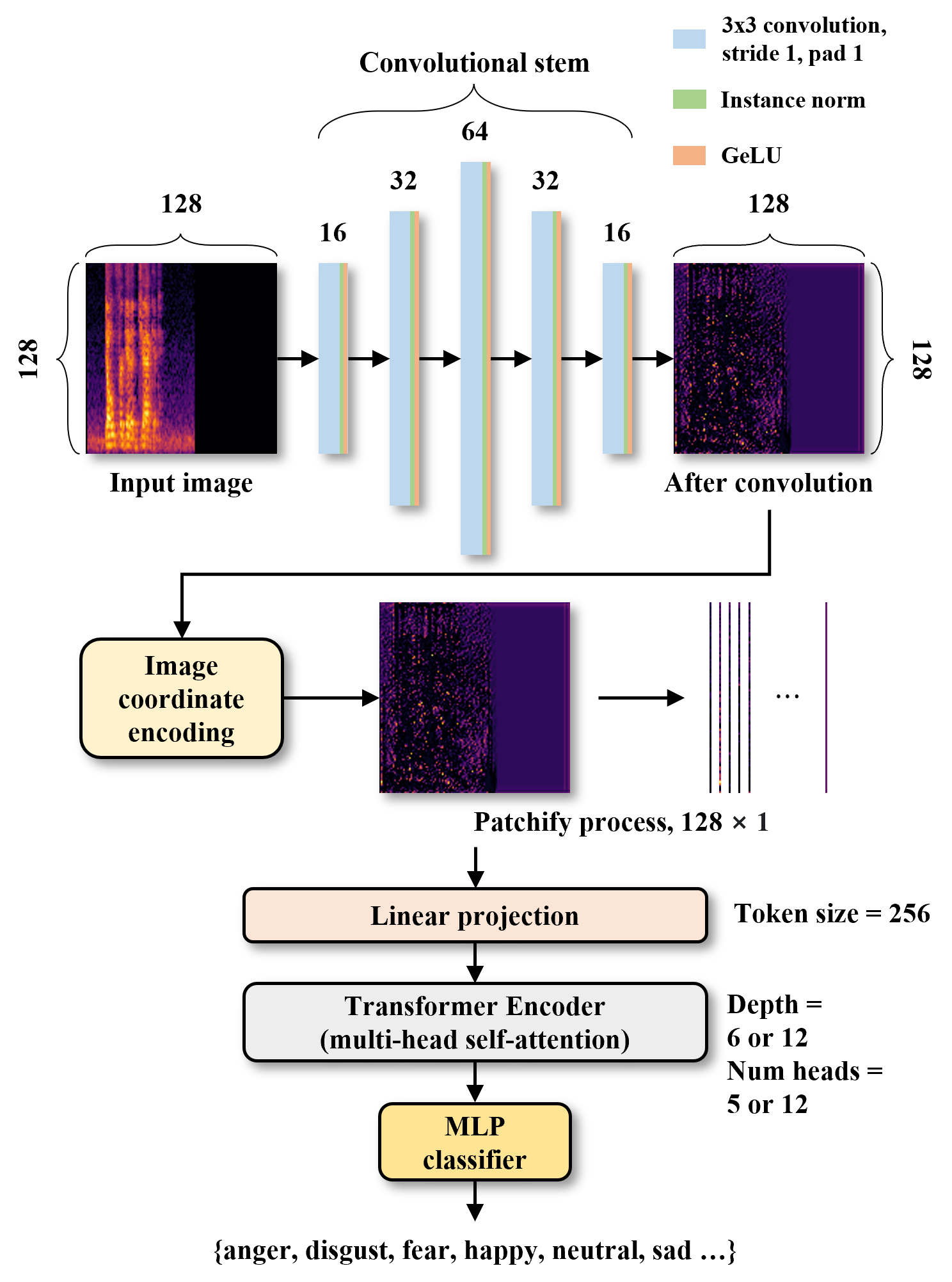}
    \caption{The structure of the teacher network. It contains a convolutional stem to obtain locality and an image coordinate encoding to disambiguate the location information.}
    \label{fig4}
\end{figure}

The student network is a ViT consisting purely of multi-head self-attention, without convolutional stem and positional encoding. The patch size is 128 x 1, which is the same as the teacher network, because the input sequence between the teacher network and the student network must be similar to extract similar features, and knowledge transfer through feature map matching and verify implementation for locality and positional encoding are possible. The student network has token size = 256, depth = 3, and num heads = 5. Figure 5 shows the structure of the student network.

\begin{figure}[t!]
    \centering
    \includegraphics[width=350pt]{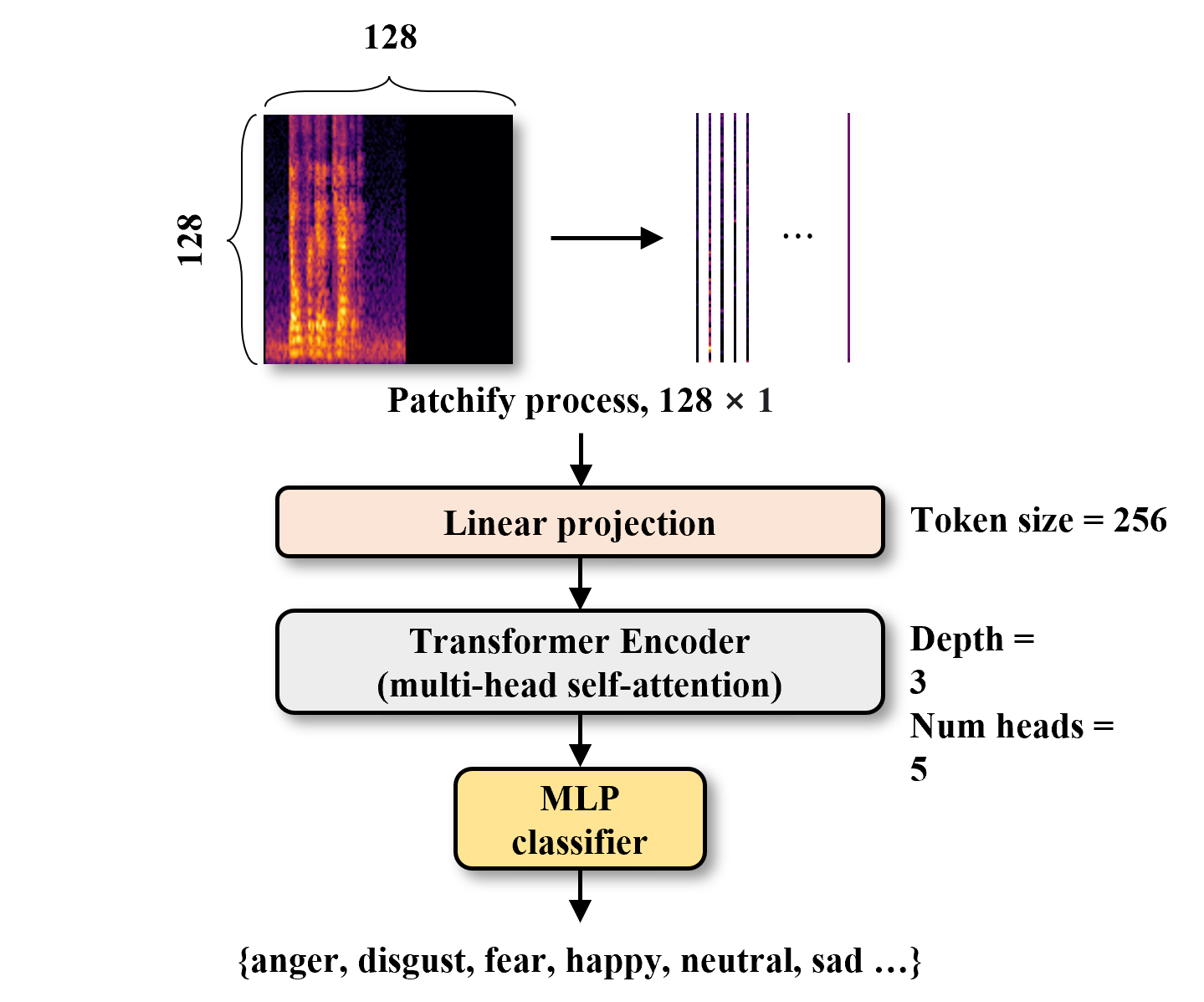}
    \caption{Structure of the student network. To verify that locality inference and positional encoding is possible through transfer learning without convolutional stem and image coordinate encoding, we construct a basic ViT.}
    \label{fig5}
\end{figure}

\subsection{Feature map matching}
We perform feature map matching to effectively transfer the knowledge of the teacher network to the student network. In the feature map matching step, after freezing the weights of the trained teacher network, we calculate the mean absolute error (L1 loss) between features to make the features of the student network similar to the features of the teacher network, and train to minimize it.

\section{Experiments}

\subsection{Datasets}

The \textbf{SAVEE} dataset\cite{HaqJackson_MachineAudition10} contains 480 acted English utterances recorded by four male actors and consists of seven emotion categories: anger, fear, disgust, happiness, neutral, sadness, and surprise.

The \textbf{EmoDB} dataset\cite{DBLP:conf/interspeech/BurkhardtPRSW05} consists of 535 German utterances from 10 actors (5 female, 5 male) and includes seven emotion categories: anger, anxiety, boredom, disgust, happiness, neutral, and sadness.

The \textbf{CREMA-D} dataset\cite{cao2014crema} consists of 7,442 videos of 91 actors (48 male and 43 female) from various ethnic groups. Actors portray a variety of emotions by uttering 12 specific sentences that correspond to one of six emotional categories: anger, fear, disgust, happiness, neutrality, and sadness.

For each dataset, the data consists of speech waveforms collected from all participants, and was divided into trainset and testset at a ratio of 80\%:20\%. A sampling frequency of 16kHz was applied to all datasets, the duration was limited to 4 seconds, and zero padding was added to the empty space of samples less than 4 seconds. Additionally, we apply STFT with N = 1024, H = 64, window size = 512 and hamming window to convert them into log-Mel spectrogram images. In this paper, we perform data augmentation of noise perturbation, time shifting, and speed perturbation. The size of the generated log-Mel spectrogram image is 128 x 128.

\subsection{Experimental setup}
In our experiments, similar to\cite{DBLP:journals/corr/abs-2203-09581}, we train with 128 x 1 patch size for teacher network, feature map matching, and student network, respectively, with hyperparameters of batch size = 4, epoch = 50, and learning rate = 1e-4 (halves every 10 epochs). It is optimized with Adam. Afterwards, we train ViT with a 16 x 16 patch size for comparative experiments with different patch shapes. Classification loss calculates cross entropy loss and feature map matching loss calculates L1 loss. We use cross entropy loss for training the teacher network, L1 loss for training the feature map matching, and cross entropy + $\alpha$ * L1 loss for training the student network. (in this case $\alpha$ = 10)

All experiments were performed on a Windows 10 workstation using an AMD Ryzen 5 7600X 6-Core Processor (4.70 GHz), 32 GB of RAM, and Nvidia RTX 4090 GPU (24GB of RAM, CUDA Cores: 16384). The entire workflow was implemented using CUDA library version 11.8 and cuDNN 8.9.3 in Pytorch version 2.0.1.

\subsection{Evaluation}
The attention mask is a visual representation of which parts of the input image are most important to the attention weight matrix as a result of performing self-attention. We first compare the difference between the attention masks of square patches and vertically elongated patches. If the attention mechanism leads to higher weighting of features that are important for the classification task, then similar attention masks should be derived for different augmented log-Mel spectrogram images. In this paper, for the attention mask comparison, we extracted attention masks from ViT using vertically elongated patches of size 128 x 1 and ViT using square patches of size 16 x 16. Both networks include convolutional stem and image coordinate encoding for comparison under the same conditions. The extracted attention masks were subjected to gaussian smoothing for ease of analysis. Figure 5 visually illustrates the experimental results of the difference in attention masks based on patch shape using randomly selected images from EmoDB. In case (a), which uses the method proposed in this paper, there is a common distinct line at the beginning and end of the signal, which indicates that similar images are recognized and attended to. However, in the case of (b), which uses a conventional square patch, we can see that it is difficult to find a common point between the attention masks. Furthermore, the time-shifted image shows confusing results with the highest attendance to the wrong zero padding region. As a result, we show that it is more robust and valid to analyze the association of frequency (y axis) with time (x axis) in a spectrogram for SER.

\begin{figure}[t!]
    \centering
    \includegraphics[width=350pt]{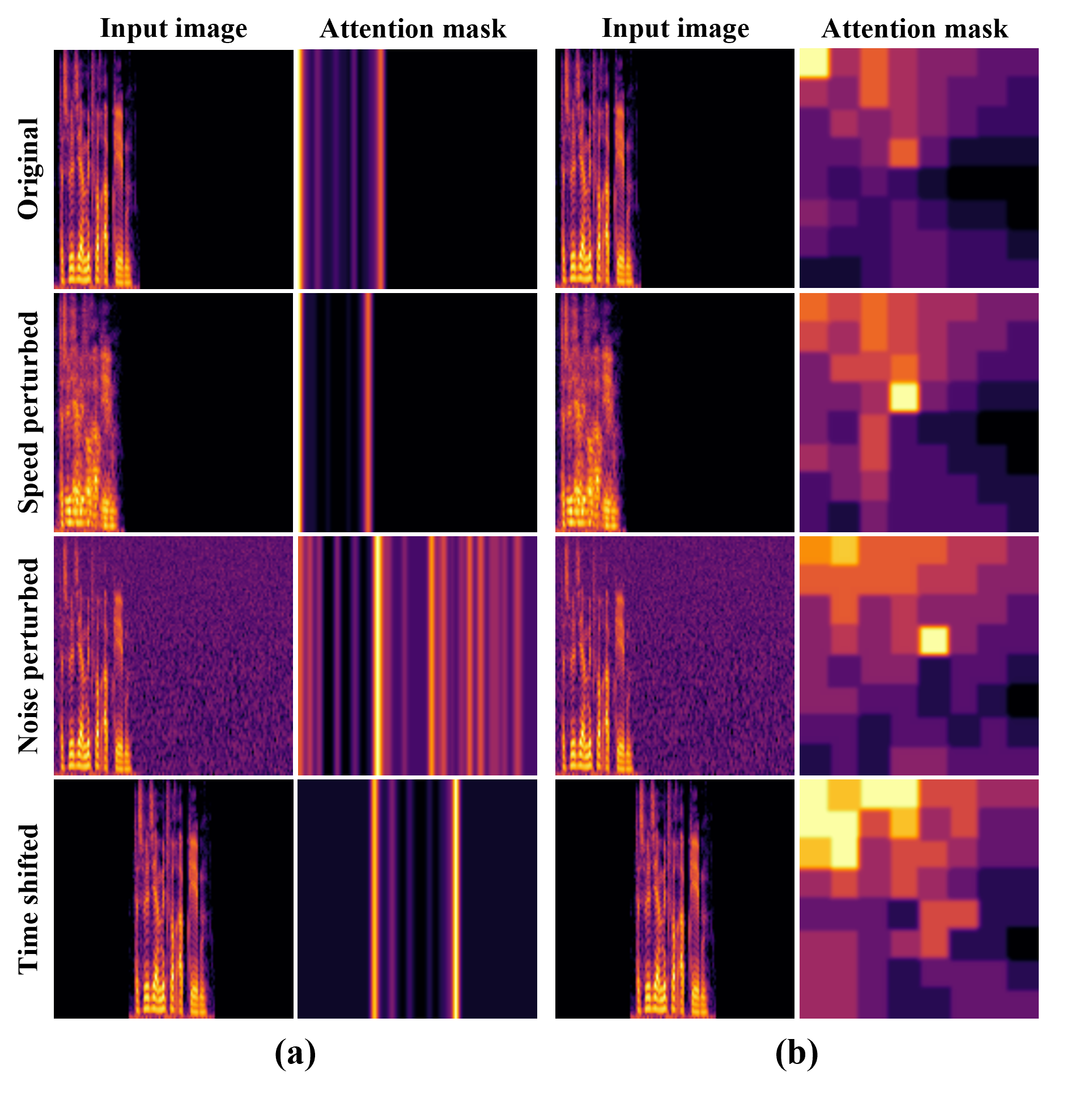}
    \caption{Comparison of attention masks from our proposed method (a) and using patches of size 16 x 16, (b). In case (a), where we analyzed the temporal frequencies correlation with vertically segmented patches, there is a common bright line (high attention weight) at the beginning and end of the signal. However, in case (b), the commonalities between the images are not well found and when time-shifted, the results are chaotic, with the highly attended in the zero padding region.}
    \label{fig6}
\end{figure}

To verify the impact of the image coordinate encoding proposed in this paper, we compared the weighted accuracy of the teacher network with depth = 6 and num heads = 5 with different positional encodings. Weighted accuracy is defined as follows:

\begin{equation}
WA=\frac{\sum\limits_{i=1}^c N_{TP_i}}{\sum\limits_{i=1}^c N_i}
\end{equation}

$N_i$ is the total number of instances evaluated and $N_{TP_i}$ is the number of instances correctly classified for class i.

In Table 1, $teacher_{no pe}$ is a teacher network that does not perform positional encoding, and $teacher_{ice}$ is a teacher network that performs image coordinate encoding. We can see that $teacher_{no pe}$ performs poorly on all three datasets. This difference is more pronounced on CREMA-D than on SAVEE and EmoDB, which are relatively small datasets. The weighted accuracy in Table 1 is the result of discarding the second decimal place.

\begin{table}[htp]
\centering
\caption{Weighted accuracy comparison between without positional encoding ($teacher_{no pe}$) vs. image coordinate encoding ($teacher_{ice}$)}
\begin{tabular}{c|c|c|c}
\hline \hline
\multicolumn{1}{c|}{\textbf{Method}} & \textbf{SAVEE↑} & \textbf{EmoDB↑} & \textbf{CREMA-D↑} \\
\hline
\multicolumn{1}{c|}{$teacher_{no pe}$} & 95.57\% & 98.83\% & 91.85\% \\
\multicolumn{1}{c|}{$teacher_{ice}$} & 97.39\% & 99.06\% & 95.07\%\\
\hline \hline
\end{tabular}
\end{table}

Additionally, to prove effectiveness of the proposed knowledge transfer method through feature map matching, we compared its classification performance with state-of-the-art methods through weighted accuracy.

To respect the author's experiment setup for AST\cite{gong21b_interspeech}, we applied ImageNet pretrain and the following conditions: patch size = 16 x 16, token size = 768, depth = 12, num heads = 12, etc. We also respected the author's experiments setup for SepTr\cite{DBLP:journals/corr/abs-2203-09581}, we applied the following conditions: patch size = 1 x 1, token size = 256, depth of vertical SepTr = 3, depth of horizontal SepTr = 3, num heads = 5, etc. As a result of the comparison in Table 2 and 3, our proposed method (student) significantly improved performance by more than 5-10\% with fewer FLOPs than the state-of-the-art methods on all datasets used for evaluation. The weighted accuracy in Table 2 and 3 is the result of discarding the second decimal place.

\begin{table}[htp]
\setlength{\tabcolsep}{5pt}
\centering
\caption{Weighted accuracy and FLOPs comparison with state-of-art-methods about depth = 12, num head = 12 networks. Student network has following condition: depth = 3, num head = 5}
\begin{tabular}{c|cccc}
\hline \hline
\multicolumn{1}{c|}{\textbf{Method}} & \textbf{SAVEE↑} & \textbf{EmoDB↑} & \textbf{CREMA-D↑} & \textbf{FLOPs↓} \\
\hline
\multicolumn{1}{c|}{Gong et al.\cite{gong21b_interspeech}} & 89.32\% & 96.02\% & 88.79\% & 3.32G \\
\multicolumn{1}{c|}{Teacher(Ours)} & 98.17\% & 96.96\% & 92.64\% & 3.58G \\
\multicolumn{1}{c|}{Student(Ours)} & \textbf{98.95\%} & \textbf{98.83\%} & \textbf{94.07\%} & \textbf{0.32G} \\
\hline \hline
\end{tabular}
\end{table}

\begin{table}[htp]
\setlength{\tabcolsep}{5pt}
\centering
\caption{Weighted accuracy and FLOPs comparison with state-of-art-methods about depth = 6, num head = 5 networks. Student network has following condition: depth = 3, num head = 5}
\begin{tabular}{c|cccc}
\hline \hline
\multicolumn{1}{c|}{\textbf{Method}} & \textbf{SAVEE↑} & \textbf{EmoDB↑} & \textbf{CREMA-D↑} & \textbf{FLOPs↓} \\
\hline
\multicolumn{1}{c|}{Ristea et al.\cite{DBLP:journals/corr/abs-2203-09581}} & 87.23\% & 92.75\% & 79.94\% & 175.81G \\
\multicolumn{1}{c|}{Teacher(Ours)} & 97.39\% & 99.06\% & 95.07\% & 1.43G \\
\multicolumn{1}{c|}{Student(Ours)} & \textbf{99.47\%} & \textbf{99.76\%} & \textbf{95.24\%} & \textbf{0.32G} \\
\hline \hline
\end{tabular}
\end{table}

\bibliographystyle{unsrt}  
\bibliography{references}  %%% Remove comment to use the external .bib file (using bibtex).


% \end{thebibliography}

\end{document}